\documentclass[10pt, a4paper]{article}
\usepackage{lrec}
\usepackage{multibib}
\newcites{languageresource}{Language Resources}
\usepackage{graphicx}
\usepackage{tabularx}
\usepackage{soul}

\usepackage{epstopdf}
\usepackage[latin1]{inputenc}

\usepackage{hyperref}
\usepackage{paralist}
\usepackage{xcolor}

\title{Bollywood Movie Corpus for Text, Images and Videos\\ \vspace*{.5\baselineskip} }

\name{Nishtha Madaan\textsuperscript{1}, Sameep Mehta\textsuperscript{1}, Mayank Saxena\textsuperscript{2}, , Aditi Aggarwal\textsuperscript{3}, Taneea S Agrawaal\textsuperscript{3}, \\
Vrinda Malhotra\textsuperscript{3}\\ }

\address{\textsuperscript{1}IBM Research-INDIA\\
\textsuperscript{2}Delhi Technological University\\
\textsuperscript{3}IIIT-Delhi\\
\{nishthamadaan, sameepmehta\}@in.ibm.com , mayank26saxena@gmail.com, \{taneea14166, vrinda14122, \\
aditi16004\}@iiitd.ac.in
}

\abstract{
 In past few years, several data-sets have been released for text and images. We present an approach to create the data-set for use in detecting and removing gender bias from text. We also include a set of challenges we have faced while creating this corpora. In this work, we have worked with movie data from Wikipedia plots and movie trailers from YouTube. Our Bollywood Movie corpus contains ~4000 movies extracted from Wikipedia and ~880 trailers extracted from YouTube which were released from 1970-2017. The corpus contains csv files with the following data about each movie - Wikipedia title of movie, cast, plot text, co-referenced plot text, soundtrack information, link to movie poster, caption of movie poster, number of males in poster, number of females in poster. In addition to that, corresponding to each cast member the following data is available - cast name, cast gender, cast verbs, cast adjectives, cast relations, cast centrality, cast mentions. We present some preliminary results on the task of bias removal which suggest that the data-set is quite useful for performing such tasks.   \\ 
\newline \Keywords{Bollywood Movie Corpus, Gender-bias detection, Gender-bias removal} }

\begin{document}

\maketitleabstract

\section{Introduction}

In past few years, several data-sets have been released which include text, images and videos \cite{ferraro2015survey}. There are several other datasets which generate automatic descriptions using images \cite{bernardi2016automatic}.  Although there has been no past dataset that can be used to study gender bias. Gender bias detection from text has been an emerging area of interest among researchers. The next step to gender bias detection is Gender Bias removal which has been gaining a lot of attention. 

Bias Detection and removal has also been investigated in social sciences and natural language processing. The work has been mainly based on some real world observations and theories. Nonetheless, there is only a little scientific work in detecting gender bias in text. There are no publicly available data-sets for this task. While there are recent works where gender bias has been studied in different walks of life \cite{soklaridis2017gender},\cite{ macnell2015s}, \cite{carnes2015effect}, \cite{terrell2017gender}, \cite{saji2016gender}, the analysis majorly involves information retrieval tasks involving a wide variety of prior work in this area. \cite{fast2016shirtless} have worked on gender stereotypes in English fiction particularly on the Online Fiction Writing Community. The work deals primarily with the analysis of how males and females behave and are described in this online fiction. Furthermore, this work also presents that males are over-represented and finds that traditional gender stereotypes are common throughout every genre in the online fiction data used for analysis. \\ Apart from this, there have been various works where Hollywood movies have been analyzed for having such gender bias present in them \cite{blog}. Similar analysis has been done on children books and music lyrics which found that men are portrayed as strong and violent, and on the other hand, women are associated with home and are considered to be gentle and less active compared to men. These studies have been very useful to know the trend but the derivation of these analyses has been done on very small data sets. In some works, gender drives the decision for being hired in corporate organizations \cite{dobbin2012corporate}. Not just hiring, it has been shown that human resource professionals' decisions on whether an employee should get a raise have also been driven by gender stereotypes by putting down female claims of raise requests. While, when it comes to consideration of opinion, views of females are weighted less as compared to those of men \cite{otterbacher2015linguistic}. On social media and dating sites, women are judged by their appearance while men are judged mostly by how they behave \cite{rose2012face}. When considering occupation, females are often designated lower level roles as compared to their male counterparts in image search results of occupations \cite{kay2015unequal}. In our work we work with Bollywood movies to create a dataset which can be leveraged for removing such biases.\\ 
The Bollywood Movie Corpus consists of data of 4000 movies extracted from Wikipedia and 880 trailers extracted from YouTube which were released from 1970-2017. In addition to this, it contains images for these 4000 movies. The dataset majorly contains co-referenced text retrieved using Wikipedia plot text and presence and absence of male and female in text, posters and trailers. This makes the corpus large and diverse enough to perform a qualitative and quantitative analysis on studying the bias present in the data. We hope that this analysis is useful to the researchers in text and image domain to also remove such bias present in the dataset and devise ways to generate bias-free stories. The main aim to publish this corpus is to enable researchers to accelerate research in this direction.

The Bollywood Movie Corpus consists of csv files with the following data about each movie - movie title, cast information, plot text, soundtrack data, poster link, poster caption, trailer link. Note that we have not included the posters and videos themselves but have added their links. We chose Wikipedia over other available options is due to the diverse and rich nature of information present in Wikipedia.

In further sections of the paper, we discuss how we have created the corpus and how it can be used.

\begin{figure*}[t!]
\includegraphics[width=1.0\linewidth]{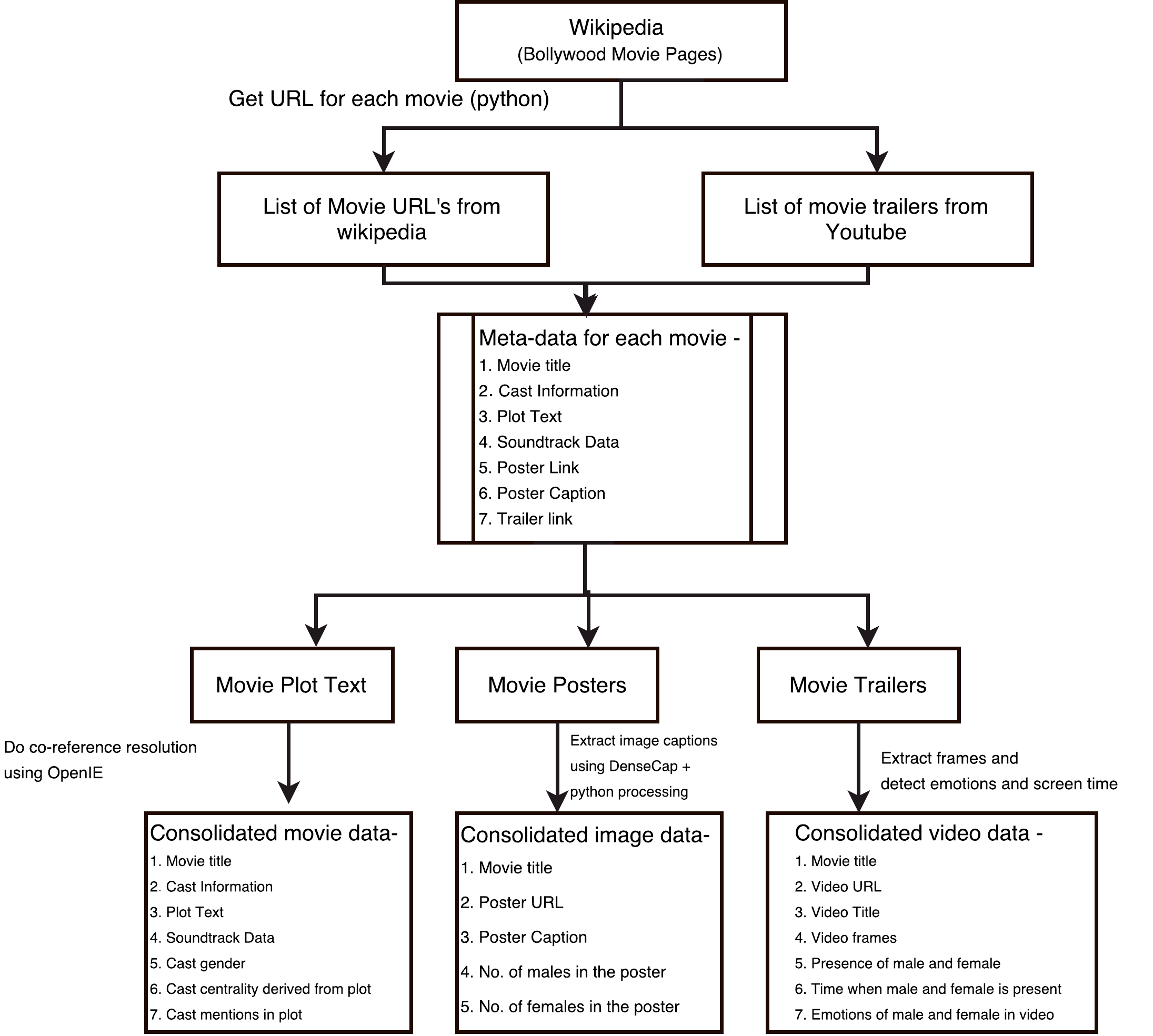}
  \caption{The Bollywood Movie Corpus Dataset Creation Process }
  \label{fig:datacreation}
\end{figure*}

\section{Dataset Creation}

Figure \ref{fig:datacreation} illustrates the steps involved in Dataset creation. We iterate one-by-one over each step and explain briefly how it has been done.

\subsection{Extraction of meta-data corresponding to each movie}
We retrieved the URL's of all movies released from 1970 to 2017. Using these URL's we extracted the information corresponding to each movie such as - movie title, movie plot text, cast information (containing information about each cast member), soundtrack information (containing information about singers of each song in the movie), links to poster in each movie. 

After the meta-data retrieval we processed the data to extract our task specific information which is presented in further subsections.

\subsubsection{Movie Plot text}
The movie plot text is processed using OpenIE. OpenIE is provided by Stanford which has been used in python. We use its co-referencing feature to generate co-referenced text plot for each movie. We use it to obtain chains of connections and then later on associate the parts of that chain to a head noun. In this way we generate co-referenced text. Also we use Stanford Dependency parser to extract relations, verbs, adjectives corresponding to each cast of the movie from the text. These are extracted by looking up the connections of cast member to the nearest adjective, verb and relation tags.

\subsubsection{Movie Posters}
The movie posters are processed using DenseCap. DenseCap is provided by Stanford which has been used to extract captions with certain confidence scores. We computed presence and absence of man and woman by analyzing the captions text generated by DenseCap. As a result, we produced for each movie poster a man is present or a woman is present or both are present.

\subsubsection{Movie Trailers Data}
The trailers dataset consists of 880 movie trailers which have been downloaded from YouTube. According to Wikipedia, there were 1026 Bollywood movies which have been released between 2008-2017. Out of these 1026 movies, we were successfully able to download official movie trailers for 880 movies i.e. for 85.7\% of the movies.
\begin{asparaenum}
\item Trailer Data Collection - For all these 1026 movies, a string was generated which was of the format - "year\textunderscore of\textunderscore release\textunderscore of\textunderscore movie" + "movie\textunderscore name" + "official\textunderscore trailer". This string was used for searching for the official movie trailer for the given movie on YouTube. The first search result was considered as the official movie trailer and the corresponding video was downloaded. The Python libraries which were used in this task were Beautiful Soup (https://pypi.python.org/pypi/beautifulsoup4) and pytube (https://pypi.python.org/pypi/pytube/6.4.2). Beautiful Soup was used for searching on YouTube and obtaining the URL of the videos. The pytube library was used for downloading the YouTube video from the URL which we had obtained earlier. Each video which was downloaded was of 480p resolution and had a frame rate of 25 frames per second and a .mp4 format.

Out of the 1026 movies, 880 of them had official trailers uploaded on YouTube which were the first result which was obtained using the search string which is mentioned above. In some cases, the full length movie was available on YouTube and was the first search result. In such cases, the full length movie was downloaded. These videos were later removed from the dataset.

\item Trailer Data Processing - Since each video had a frame rate of 25 frames per second, we obtained every 25th frame of the video using OpenCV (https://pypi.python.org/pypi/opencv-python) library. Now, we had a frame for every second of the video. The next step which was involved in the processing of the data was to detect the presence/absence of a person in the frame. And if a person was detected, the final step was to detect the gender of the person and the emotion exhibited by the person in the frame. This procedure was followed for all the videos in the dataset and all the results were saved in a corresponding .csv files for each video. 

The tool which was used to run gender and emotion detection is an open source tool. It has been developed by researchers at Hochschule Bonn Rhein Sieg University.(https://github.com/oarriaga/face\textunderscore classification). The tool has a MIT license and is available for commercial as well as private use.

Now, each frame in which a person was detected had 2 labels - (I) Gender of person (II) Emotion displayed by person. Using this information, the data-set of labeled frames was further dis-aggregated into different categories. Some of them being - (I) Gender (II) Gender vs Emotion (III) Emotion.
\end{asparaenum}

\section{Expected Applications}
\begin{enumerate}
    \item In creation of unbiased stories - This dataset can be used to generate unbiased plausible stories from biased stories. The main area where this can be extended is to train classifiers to identify which is a biased statement and which is not. The information in data supplied in our Corpus can be used as a surrogate for biased information. Using this information one can come up with a model which takes this data and feeds certain features of this text into the model and generate bias free stories. 
    \item To identify bias in other pieces of text - Our dataset is once corpus which highlights what bias is present at sentence and inter sentence level. This information can be used to identify similar biases in other contexts. Authors believe since there was a dier need of such a dataset in identifying biases in our current systems, this is the first dataset which can be treated as a benchmark dataset for such works in future.
    \item To create a fact data for bias - Our dataset can be used to gather fact dataset since it makes it easier to identify pieces of information in text to be biased or not. This opens up new possibilities of generating a fact dataset of our own which is free from any such bias. This work can be taken as motivation to such work.
\end{enumerate}

\section{Availability Format}

The Bollywood Movie Corpus is published at Github in a private repository as a set of csv files which can be opened in MS excel. This is accompanied by python scripts for crunching text, image and videos data. The schema details have been provided in README file accompanying these scripts. 
The scripts and data will facilitate a user to be able to add new data files and generate similar outputs on text and images as we have obtained for Bollywood Movie Corpus. We will make the data in Github public once we get acceptance on the work.

\section{Conclusion}
In this paper, we introduced a Bollywood Movie Corpus containing text, images and videos. The corpus contains over 4000 Bollywood movies data. We hope that this data can be used to devise algorithms to generate bias free stories that can benefit future generations by learning how things are rather than inheriting bias from past stories.

\bibliographystyle{lrec}
\bibliography{xample}

\bibliographystylelanguageresource{lrec}
\bibliographylanguageresource{xample}

\end{document}